\documentclass[twocolumn,showpacs,preprintnumbers,amsmath,amssymb,showkeys]{revtex4}
\usepackage{amsfonts}
\usepackage{graphicx, epsfig, amsmath}
\usepackage{dcolumn}
\usepackage{bm}
\usepackage{epstopdf}
\usepackage{hyperref}

\newcommand{\bo}{\raise-1mm\hbox{\Large$\Box$}}

\newcommand{\f}[2]{\frac{#1}{#2}}

\newcommand{\w}{\omega}
\newcommand{\kp}{\kappa}
\newcommand{\be}{\begin{equation}}
\newcommand{\ee}{\end{equation}}
\newcommand{\bea}{\begin{eqnarray}}
\newcommand{\eea}{\end{eqnarray}}
\newcommand{\eqn}[1]{(\ref{#1})}

\begin{document}



\title{On Spin-Statistics and Bogoliubov Transformations in Flat Spacetime With Acceleration Conditions}





\author{Michael R.R. Good}
	\email{mgood@ntu.edu.sg}
	\affiliation{Institute of Advanced Studies, Singapore}
\date{\today}

\begin{abstract}
A single real scalar field of spin zero obeying the Klein-Gordon equation in flat spacetime under external conditions is considered in the context of the spin-statistics connection.  An imposed accelerated boundary on the field is made to become, in the far future, (1) asymptotically inertial and (2) asymptotically non-inertial (with an infinite acceleration).  The constant acceleration Unruh effect is also considered.  The systems involving non-trivial Bogoliubov transformations contain dynamics which point to commutation relations.  Particles described by in-modes obey the same statistics as particles described by out-modes.  It is found in the non-trivial systems that the spin-statistics connection can be manifest from the acceleration.  The equation of motion for the boundary which forever emits thermal radiation is revealed.
\end{abstract}
\maketitle

\section{INTRODUCTION}

Parker demonstrated that the spin-statistics connection is evident from the dynamics of curved spacetime.\cite{Parker:1969au} \cite{Parker2} \cite{statdyn}  It was first shown using a quantized real scalar field of spin zero which obeys the curved spacetime Klein-Gordon equation\cite{Parker:1969au}.  It was then found that consistency resulted when commutation relations imposed on the creation and annihilation operators for the in-region, were also imposed for the creation and annihilation operators for the out-region.  It is not consistent with the dynamics to impose anti-commutation relations for the out-region. 

%

%
Parker's derivation was extended to fermions of spin-1/2 which obey the Dirac equation and satisfy anticommutation relations \cite{Parker2}. Higher spin and parastatistics was treated \cite{statdyn}, the statistics were generalized \cite{nogenstat}, \cite{Bardek:1994wh}, \cite{Goodison:1995sk}, and ghost fields \cite{Parkerghost} were examined. It is assumed \cite{Parkerbook} that a connection between statistics and dynamics is not present in Minkowski spacetime.  It is also generally accepted that particles at late times should obey the same statistics as at early times.

Recently, researchers may have presented the first experimental evidence for observation of the dynamical Casimir effect \cite{BOOYAA}. Within the context of an imposed boundary in flat spacetime, the question arises whether the connection between spin and statistics is revealed from the dynamics of a field-boundary system in the same way as from the dynamics of curved spacetime.  

In this note, it is shown that if a quantized real scalar field of spin-0 obeys the flat spacetime Klein-Gordon equation with an imposed accelerated boundary condition, then commutation relations are the admissible algebra that may be imposed upon the creation and annihilation operators on the in-region and the out-region.  It will not be consistent with the dynamics to establish anti-commutation relations upon this field-boundary system.

As is pointed out by Wald \cite{Wald:1979wt}, the spin-statistics results from curved spacetime depend on the conserved inner product.  Wald emphasized that no reasonable quantum field theory in curved spacetime (or in flat spacetime with an external potential), can have Bose-Einstein statistics if the inner product is positive definite for all positive and negative frequency solutions.  Likewise, it's unreasonable to have Fermi-Dirac statistics if the inner product is positive definite only on positive frequency solutions.  In this note, it is emphasized that acceleration induces the appearance of distinct mode solutions with their respective positive and negative frequency pieces.  These distinct modes give rise to a non-trivial Bogoliubov transformation which dictates the commutation relations.  In the following three sections, it is shown that an acceleration ultimately gives rise to the spin-statistics connection in three salient examples: an asymptotically inertial mirror in Section \ref{sec:AIM} (the most simple case), an infinitely accelerated mirror in Section \ref{sec:IAM} and the Unruh effect in Section \ref{sec:Unruh}.

\section{Dynamics to Statistics}

\subsection{ Asymptotically Inertial Mirror} \label{sec:AIM}

\begin{figure}[ht]
\begin{center}
\includegraphics[scale=0.8]{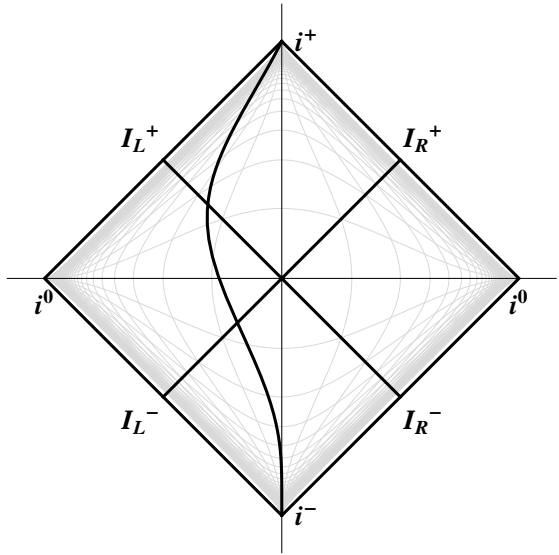}
\caption{\label{fig:Penrose}An asymptotically inertial trajectory in a Penrose diagram.  In this example, $z(t) = - \frac{1}{2}\sinh^{-1}(e^t)$.}
\end{center}
\end{figure}

\begin{figure}[ht]
\begin{center}
\rotatebox{90}{\includegraphics[scale=0.8]{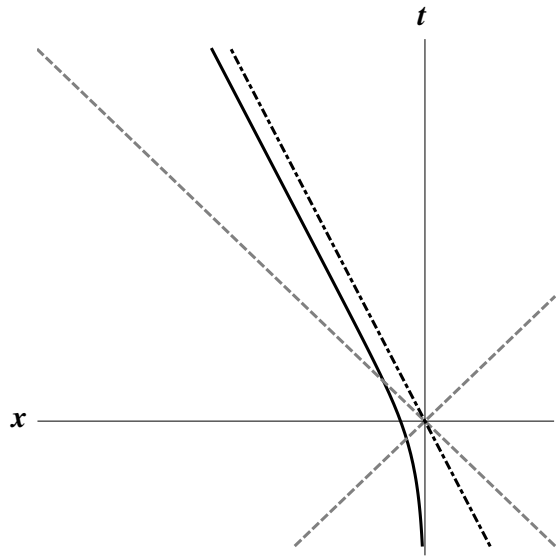}}
\caption{\label{fig:spacetime} An asymptotically inertial trajectory with a final coasting speed of half the speed of light displayed in the usual spacetime diagram.  The gray dashed lines represent the light cone, and the black dotted-dashed line shows the asymptote of the mirror trajectory.  The trajectory example here is the same as Figure \ref{fig:Penrose}. }
\end{center}
\end{figure}

The moving mirror model is the most simple example of the dynamical Casimir effect \cite{Davies:1976hi} \cite{Davies:1977yv}.  Consider a moving mirror (an external boundary condition in 1+1 dimensions) which does not accelerate forever.  This mirror starts (ends) at time-like  past (future) infinity, asymptotically having zero acceleration in the far past (future).  As such, an asymptotically inertial mirror will contain no horizons or pathological acceleration singularities, as seen in Figure \ref{fig:Penrose} and Figure \ref{fig:spacetime}.  The use of an asymptotically inertial mirror is chosen for clarity of derivation.  Relaxing the condition of asymptotically inertial motion will alter the form of the derivation and the complexity of the relevant identities.  The connection between acceleration and the spin-statistics connection remains.   The use of the forever accelerating observer in the Unruh effect, or a forever accelerating mirror that results in a null horizon, requires the appropriate left-right construction (for example, \cite{Carlitz:1986nh}) to incorporate complete Cauchy surface information. \\
  
A moving mirror in flat spacetime reveals a connection between statistics and dynamics that is not present in Minkowski spacetime without acceleration.  Statistics can be derived from the dynamics of the moving mirror as an accelerated mirror gives a beta Bogoliubov transformation coefficient which is non-zero.  The statistics are determined by the algebra of the creation operators.  The associated relations of the creation operators are commuting for Bose-Einstein statistics and anticommuting for Fermi-Dirac statistics.  The following derivation shows that for the spin-zero field, only Bose-Einstein statistics are invariable with the flat spacetime dynamics that include an accelerated external boundary condition. \\

The field equation of motion is the Klein-Gordon equation $\Box \psi = 0$.  The moving mirror is imposed such that at the position of the mirror the field is zero, $\psi|_z =0$.  This system has modes that can be used to expand the field,
\be \psi = \int_0^\infty d\w'\; \left[a_{\w'} \phi_{\w'} + a^\dagger_{\w'}\phi^*_{\w'}\right], \ee
\be \psi = \int_0^\infty d\w\; \left[b_\w \chi_\w + b^{\dagger}_\w\chi^{*}_\w\right]. \ee
The modes are orthonormal and complete.  They take the null coordinate form,
\be 
\phi_{\w'} = (4\pi \w')^{-1/2} [e^{-i\w' v } - e^{-i\w' p(u)} ],
\ee
\be
\chi_{\w}  = (4\pi \w)^{-1/2} [e^{-i\w f(v)} - e^{-i\w u} ].
\ee
The modes can be expanded in terms of each other, 
\be \label{phichi} \phi_{\w'} = \int_0^\infty d\w\; \left[\alpha_{\w'\w} \chi_{\w} + \beta_{\w'\w}\chi^*_{\w}\right], \ee
\be \label{chiphi} \chi_{\w} = \int_0^\infty d\w'\; \left[\alpha^*_{\w'\w} \phi_{\w'} - \beta_{\w'\w}\phi^*_{\w'}\right]. \ee
by introduction of Bogoliubov coefficients,  
\be \alpha_{\w'\w} = (\phi_{\w'},\chi_{\w}), \qquad \beta_{\w'\w} = -(\phi_{\w'},\chi^*_{\w}), \ee
defined by the flat-space scalar product, in spacetime coordinates, 
\be (\phi_{\w'} , \chi_{\w} ) = i \int_{-\infty}^{\infty} dx\; \phi_{\w'}^* \stackrel{\leftrightarrow}{\partial_t}\chi_{\w}, \ee
or in null coordinates,
\be (\phi_{\w'} , \chi_{\w} ) = i \int_{-\infty}^{\infty} du\; \phi_{\w'}^* \stackrel{\leftrightarrow}{\partial_u} \chi_{\w} + i \int_{-\infty}^{\infty} dv\; \phi_{\w'}^* \stackrel{\leftrightarrow}{\partial_v} \chi_{\w}. \ee
The field equation combined with the moving mirror boundary condition imply that the Bogoliubov coefficients $\alpha_{\w'\w}$ and $\beta_{\w'\w}$ relate the operators $a_{\w'}$ and $a_{\w'}^\dagger$ to the operators $b_{\w}$ and $b^{\dagger}_{\w}$:
\be \label{aforBog} a_{\w'} =  \int d\w \left[\alpha^*_{\w'\w} b_{\w} - \beta^{*}_{\w'\w} b^{\dagger}_{\w}\right], \ee
\be \label{bforBog} b_{\w} =  \int d\w' \left[\alpha_{\w'\w} a_{\w'} + \beta^{*}_{\w'\w} a^{\dagger}_{\w'}\right]. \ee
%
The modes are orthonormal such that,
\be \label{phinormal}(\phi_{\w},\phi_{\w'}) = -(\phi^*_{\w},\phi^*_{\w'}) = \delta(\w-\w') ,\; (\phi_{\w},\phi^*_{\w'}) = 0, \ee
\be \label{chinormal}(\chi_{\w},\chi_{\w'}) = -(\chi^*_{\w},\chi^*_{\w'}) = \delta(\w-\w') ,\; (\chi_{\w},\chi^*_{\w'}) = 0. \ee

Since the modes are related by a Bogoliubov transformation, the above scalar products imply several identities. Using \eqn{phinormal}, \eqn{phichi}, and \eqn{chinormal}  the following Wronskian identity holds, 
\be\label{MGdidWron}   \int d\w'' \left[\alpha_{\w \w''} \alpha^*_{\w'\w''} - \beta_{\w \w''}\beta^{*}_{\w'\w''}\right] = \delta(\w - \w'). \ee
Note that this relation is a derived consequence only of the field equation and the imposed asymptotically inertial mirror.  This relation is a result of the dynamics.  \\

Consider now, the possible commutation relations of the creation operators, $a_{\w'}$, $a^\dagger_{\w'}$ and $b_\w$, $b^\dagger_\w$ (which are associated with the $\phi_{\w'}$ and $\chi_{\w}$ modes, respectively),
\be \label{aposs} [a_\w , a^\dagger_{\w'} ]_{\pm} = \delta(\w'-\w) ,\;[a_\w , a_{\w'} ]_{\pm} = [a^\dagger_\w , a^\dagger_{\w'} ]_{\pm} = 0, \ee
\be \label{bposs} [b_\w , b^\dagger_{\w'} ]_{\pm} = \delta(\w'-\w) ,\; [b_\w , b_{\w'} ]_{\pm} = [b^\dagger_\w , b^\dagger_{\w'} ]_{\pm} = 0, \ee
Here the $+$ sign corresponds to Fermi-Dirac statistics from the anticommutator, while the $-$ sign corresponds to Bose-Einstein statistics from the commutator.  The relations associated with the $\phi_{\w'}$ modes may be expressed in terms of Bogoliubov coefficients using \eqn{aforBog},
\be \label{MGdidit1} [a_\w , a^{\dagger}_{\w'} ]_{\pm} = \int d\w'' \left[\alpha_{\w''\w}\alpha^*_{\w''\w'} \pm \beta_{\w''\w'}\beta^{*}_{\w''\w}\right]. \ee
The relations associated with the $\chi_{\w}$ modes may be expressed in terms of Bogoliubov coefficients using \eqn{bforBog},
\be \label{MGdidit2} [b_\w , b^{\dagger}_{\w'} ]_{\pm} = \int d\w'' \left[\alpha_{\w''\w}\alpha^*_{\w''\w'} \pm \beta_{\w''\w'}\beta^{*}_{\w''\w}\right]. \ee
The dynamics-only result \eqn{MGdidWron} dictates which sign to use in \eqn{MGdidit1} and \eqn{MGdidit2}.  Since the dynamics indicates the choice of the $-$ sign, the election of Bose-Einstein statistics is straightforward. Similar identities to \eqn{MGdidWron}, using \eqn{phinormal} and \eqn{chinormal}, dictate the correct sign to be used in \eqn{aposs} and \eqn{bposs} which give, in total,
\be [a_\w , a^\dagger_{\w'} ]_{-} = \delta(\w'-\w),\; [a_\w , a_{\w'} ]_{-} = [a^\dagger_\w , a^\dagger_{\w'} ]_{-} = 0, \ee
\be [b_\w , b^\dagger_{\w'} ]_{-} = \delta(\w'-\w),\; [b_\w , b_{\w'} ]_{-} = [b^\dagger_\w , b^\dagger_{\w'} ]_{-} = 0. \ee
Particles at different times, (those associated with $b^\dagger_{\w}b_{\w}$ and $a^\dagger_{\w'}a_{\w'}$) obey the same statistics only in the Bose-Einstein case.  The Bose-Einstein case is then fixed with the dynamics of the field and mirror.  When the mirror has acceleration, the $\beta$ Bogoliubov transformation coefficient is not zero.  If the mirror is not accelerating, then $\beta = 0$, and this connection between dynamics and statistics is absent, despite the presence of a boundary and the abandonment of Minkowski spacetime.  The spin-statistics-dynamics connection in flat spacetime relies on a non-zero acceleration.  The derivation presented above is an `accelerated boundary condition in flat spacetime' \textendash{}moving mirror\textendash{} derivation of the relationship between spin and statistics.

\subsection{  Infinitely Accelerated Mirror} \label{sec:IAM}
\begin{figure}[ht]
\begin{center}
\includegraphics[scale=0.8]{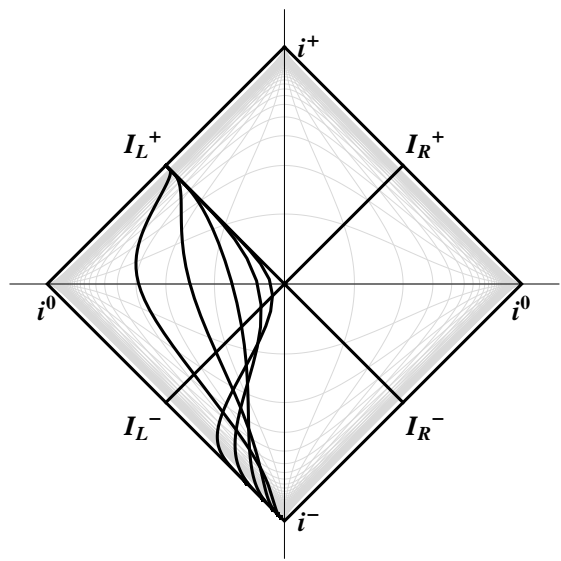}
\caption{\label{fig:PenroseIAM}Asymptotically infinite accelerating trajectories in a Penrose diagram. In this figure, all mirrors emit thermal radiation at all times with $T=\kappa/2\pi$ along the trajectory $z(t) = -t- \frac{1}{\kappa}W(e^{-2\kappa t})$, where $\kappa = \pi/8,\; \pi/4,\; \pi/2,\; \pi,\; 2\pi$.}
\end{center}
\end{figure}

\begin{figure}[ht]
\begin{center}
\rotatebox{90}{\includegraphics[scale=0.8]{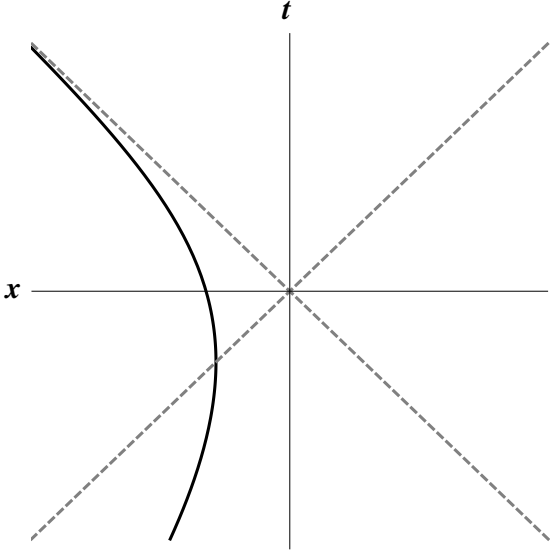}}
\caption{\label{fig:SpacetimeIAM}The asymptotically infinite accelerating trajectory of the Carlitz-Willey mirror in a spacetime diagram. The mirror emits thermal radiation at all times with $T=\kappa/2\pi$ along the trajectory $z(t) = -t- \frac{1}{\kappa}W(e^{-2\kappa t})$. Notice the Carlitz-Willey mirror does not start statically, and also approaches the speed of light in the future with an acceleration singularity, resulting in a horizon.}
\end{center}
\end{figure}

Consider a moving mirror which accelerates for all time.  When a mirror in flat spacetime accelerates forever, a horizon singularity is formed. The horizon can be seen most easily in a Penrose diagram, (see Figure \ref{fig:PenroseIAM} or in a spacetime diagram in Figure \ref{fig:SpacetimeIAM}). It is reasonable to ask if the physically pathological singularity breaks the dynamics-statistics connection.  As it turns out, the acceleration singularity complicates the mathematics but does not ruin the dynamics-statistics link.  The spin-statistics connection is apparent in the case of an infinitely accelerated boundary. 

In this section only, we adopt Carlitz-Willey's normalization, transformation conventions and left-right construction \cite{Carlitz:1986nh}.  The advantage here is the simplicity of the explicit mode forms.  The field motion is the Klein-Gordon equation $\Box \psi(u,v) = 0 $, and the field is written as

\be \psi = \f{1}{4\pi}\int_0^\infty \f{d\w'}{\w'}\left[a_{\w'} \phi_{\w'} + a^\dagger_{\w'}\phi^*_{\w'}\right] \ee

\be \psi = \f{1}{4\pi}\sum_I\int_0^\infty \f{d\w}{\w}\left[b^I_\w \chi^I_\w + b^{I \dagger}_\w\chi^{I*}_\w\right] \ee
where $I =R,L$.  The modes take the simplified form,

\be \phi_{\w'} = e^{-i\w' v } - e^{-i\w' p(u)} 
\ee
\be \left\{ \begin{array}{l l}
 \chi^R_{\w}  = e^{-i\w f_R(v)}\Theta(-v) - e^{-i\w u} \\
 \chi^L_{\w}  = e^{-i\w f_L(v)}\Theta(v) \\ 
  \end{array} \right.
 \ee
while the ray-tracing functions for the specified Carlitz-Willey trajectory assume 
\be p(u) = - \f{e^{-\kp u}}{\kp} \label{p-carlitz-willey}\ee
\be \left\{ \begin{array}{l l}
 f_R(v) = -\kp^{-1}\ln(-\kp v) \quad v<0\\
 f_L(v) = +\kp^{-1}\ln(+\kp v) \quad v>0.\\ 
  \end{array} \right.
 \ee
Carlitz and Willey point out that a constant energy flux emitted by the mirror, $\frac{\kappa^2}{48 \pi}$, is obtained by the ray tracing function of \eqn{p-carlitz-willey}.  Substituting $p(u) = t + z(t)$ into \eqn{p-carlitz-willey} gives the resulting condition on the trajectory,
\be t + z(t) = - \frac{1}{\kappa} e^{-\kappa t  + \kappa z(t)}  \;. \ee
Carlitz and Willey did not provide an analytic solution to this equation.  However one can be obtained using the Lambert $W$ function (the product logarithm) with the result that
\be z(t) = -t - \f{1}{\kp}W(e^{-2\kappa t})  \;.\ee
Figure \ref{fig:PenroseIAM} and Figure \ref{fig:SpacetimeIAM} are plots of this trajectory.  It is straightforward to show that $\dot{z}\rightarrow \mp 1$ in the limits $t \rightarrow \pm \infty$ and that $z < 0$ for all time.  The proper acceleration is
\be \label{thermalacc} \alpha(t)= -\f{\kp}{2\sqrt{W(e^{-2 \kp t})}} \;. \ee
Note that the proper acceleration is not constant even though the energy flux is.  The acceleration \ref{thermalacc} is required to have thermal emission, for all times, in the moving mirror model.  Contrast this to the Unruh effect, where time-independent acceleration, $\kappa$, is responsible for thermal emission.  

The dynamics of the field-mirror system give rise to the Bogoliubov coefficients $\alpha^I_{\w'\w}$ and $\beta^I_{\w'\w}$ which relate the operators $a_{\w'}$ and $a_{\w'}^\dagger$ to the operators $b^I_{\w}$ and $b^{I\dagger}_{\w}$:

\be a_{\w'} = \f{1}{4\pi}\sum_I \int \f{d\w}{\w} \left[\alpha^I_{\w'\w} b^I_{\w} + \beta^{I}_{\w'\w} b^{I\dagger}_{\w}\right] \ee


Since the operators are related by a Bogoliubov transformation, it now pays to examine the the conserved scalar products for the two distinct modes $\phi$ and $\chi$, (here $J=L,R$).  Specifically,
\be  i\int_{-\infty}^{\infty} dv \phi^*_{\w'} \stackrel{\leftrightarrow}{\partial_v} \phi_{\w} = 4\pi\w\delta(\w-\w') \ee

\be i\int_0^{\infty}dv \chi^{I*}_{\w'} \stackrel{\leftrightarrow}{\partial_v} \chi^J_\w + i\int_{-\infty}^{\infty} du \chi^{I*}_{\w'} \stackrel{\leftrightarrow}{\partial_u} \chi^J_{\w} = 4\pi \w \delta(\w - \w')\delta^{IJ} \ee

The conserved scalar products imply several identities, as first shown by \cite{Carlitz:1986nh}, in particular:

\be \f{1}{4\pi} \sum_I \int \f{d\w}{\w} \left[ \alpha^{I}_{\w'\w} \alpha^{I*}_{\w''\w} - \beta^I_{\w'\w}\beta^{I*}_{\w''\w}\right] = 4\pi \w' \delta(\w'' - \w') \ee

\be \f{1}{4\pi} \sum_I \int \f{d\w}{\w} \left[ \alpha^I_{\w'\w}\beta^I_{\w''\w} - \beta^I_{\w'\w}\alpha^I_{\w''\w}\right] = 0 \ee

\be\label{btran1}  \f{1}{4\pi} \int \f{d\w'}{\w'} \left[ \alpha^{I*}_{\w'\w} \alpha^{J}_{\w'\w''} - \beta^I_{\w'\w}\beta^{J*}_{\w'\w''}\right] = 4\pi \w \delta(\w - \w'')\delta^{IJ} \ee

\be \f{1}{4\pi} \int \f{d\w'}{\w'} \left[ \alpha^{I*}_{\w'\w}\beta^J_{\w'\w''} - \beta^I_{\w'\w}\alpha^{J*}_{\w\w''}\right] = 0 \ee

These relations are derived consequences of the field equation and the imposed Carlitz-Willey mirror.  Now consider the possible commutations relations
\be [a_\w , a^\dagger_{\w'} ]_{\pm} = 4\pi \w\delta(\w-\w') \ee
\be [a_\w , a_{\w'} ]_{\pm} = [a^\dagger_\w , a^\dagger_{\w'} ]_{\pm} = 0 \ee
As before, the $+$ sign corresponds to FD statistics, while the $-$ sign corresponds to BE statistics. The possible commutation relations in terms of the transformation coefficients are

\be \label{comm1} [a_\w , a^{\dagger}_{\w'} ]_{\pm} =\f{1}{4\pi} \int \f{d\w''}{\w''} \left[ \alpha^{I*}_{\w''\w} \alpha^{J}_{\w''\w'} - \beta^I_{\w''\w}\beta^{J*}_{\w''\w'}\right]
\ee

The dynamics result \eqn{btran1} can be used to determine which of the possible commutation relations are to be used in \eqn{comm1}.  The dynamics dictates the commutation relations use the $-$ sign,
\be [a_\w , a^{\dagger}_{\w'} ]_{-} = 4\pi \w \delta(\w-\w') \ee

The acceleration singularity does not corrupt the spin-statistics-dynamics connection.  

\subsection{ Unruh Effect} \label{sec:Unruh}

\begin{figure}[ht]
\begin{center}
\includegraphics[scale=0.8]{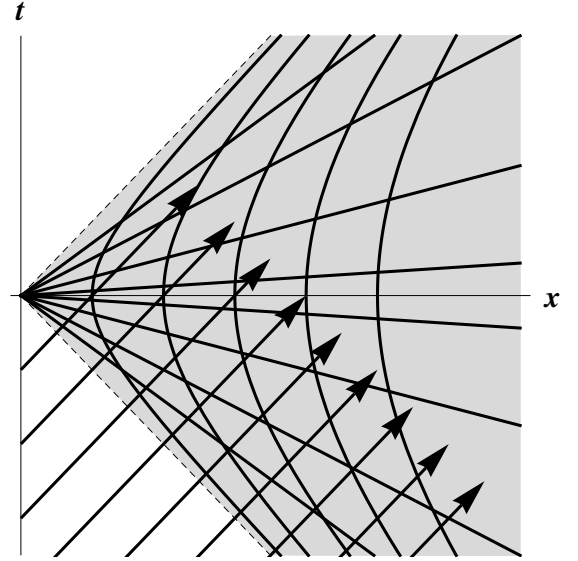}
\caption{\label{fig:RT} Right movers in the right-quadrant Rindler wedge. In this shaded spacetime region, $U<0$, and $V>0$.  The hyperbolas are constant $\tau$ values, while the radial lines are constant $\rho$ values. }
\end{center}
\end{figure}

Consider the spin-statistics connection in the setting of the Unruh effect. We specify to the right movers in the right quadrant of the Rindler wedge, $x>0$ and $x>|t|$.  The left movers follow suit. Further consideration to the left Rindler wedge is similar to the right Rindler wedge \cite{Brout:1995rd}.  To satisfy completeness on the full Cauchy surface, both the left and right movers, and both the left and right Rindler wedges must be treated.  It is found that correct commutations relations are given due to the presence of acceleration for the Rindler observer.  We start with the Minkowski metric, $ds^2 = -dt^2 + dx^2$, and Klein-Gordon field equation $(-\partial_t^2 + \partial_x^2)\psi = 0$.  Transformation to Rindler coordinates,
\be t = \rho \sinh \kappa \tau ,\quad x = \rho \cosh \kappa \tau \ee
where $\rho > 0$ and $ -\infty <\tau < \infty $, gives the metric and field equation,

\be ds^2 = -\rho^2 \kappa^2 d\tau^2 + d\rho^2 , \ee

\be \left[-\frac{1}{\kappa^2}\frac{\partial^2}{(\partial \tau)^2} + \frac{\partial^2}{[\partial(\ln \kappa \rho)]^2}\right]\psi = 0. \ee
The mode functions that are right-movers, (dependent only on $U$) and the right quadrant, R, are shown graphically in Figure \ref{fig:RT}.  The left-movers are dependent on $V$ and follow the analogous forthcoming procedure.
Specifying to these right-moving eigenmodes and the R spacetime region is clarifying.  This is done by the use of null coordinates,
\be \left. \begin{array}{l l}
 u \\
 v \\ 
 \end{array} \right\} = \tau \mp \frac{\ln \kappa \rho}{\kappa}, \ee
and
 \be \left. \begin{array}{l l}
 U \\
 V \\ 
 \end{array} \right\} = t \mp x =  \left\{ \begin{array}{l l}
 -\kappa^{-1}e^{-\kappa u} \\
 \kappa^{-1}e^{\kappa v} \\ 
 \end{array} \right. . \ee
Consider now, the two representations of the field:
\be \psi(U) =\psi^R(U) = \int_0^\infty d\omega' [ a_{\omega'} \phi_{\omega'} + a_{\omega'}^\dagger \phi_{\omega'}^*] \ee
and
\be \psi^R(U) = \theta(-U) \int_0^\infty d\omega [ b_{\omega}^R \chi_{\omega}^R + b_{\omega}^{R \dagger } \chi_{\omega}^{R *}] \ee
Notice that we have specified to the right quadrant by the use of $\theta(-U)$ and restricted our concern to right movers by utilizing the single $U$ coordinate dependence.  The advantage here is that left and right movers in the null coordinates $U$ and $V$ do not mix under the Bogoliubov transformation.   

The modes take the form,
\be \phi_{\omega'} = \frac{1}{\sqrt{4\pi \omega'}} e^{-i \omega' U}, \ee
\be \chi^R_{\omega} = \theta(-U) \frac{1}{\sqrt{4\pi \omega}}(-\kappa U)^{\frac{i\omega}{\kappa}}. \ee
The Bogoliubov transformation for the operator associated with the $\phi_{\omega'}$ takes the R-form, 
\be a_{\omega'} = \int_0^\infty dw [   \alpha^{R *}_{\omega'\omega} b^R_{\omega} - \beta_{\omega'\omega}^{R *} b^{R \dagger}_{\omega}] \ee
and the operator associated with the $\chi^R_{\omega}$ is
\be b_{\omega}^R = \int_0^\infty dw' [  \alpha_{\omega'\omega}^R a_{\omega'} + \beta_{\omega'\omega}^{R *} a_{\omega'}^\dagger] \ee
The modes are non-trivially expanded in terms of each other, 
\be \label{phichi2} \phi_{\w'} = \int_0^\infty d\w\; \left[\alpha^{R}_{\w'\w} \chi^R_{\w} + \beta^R_{\w'\w}\chi^{R*}_{\w}\right], \ee
\be \label{chiphi2} \chi^R_{\w} = \int_0^\infty d\w'\; \left[\alpha^{R*}_{\w'\w} \phi_{\w'} - \beta^R_{\w'\w}\phi^{*}_{\w'}\right]. \ee
by introduction of Bogoliubov coefficients,  
\be \alpha^R_{\w'\w} = (\phi_{\w'},\chi^R_{\w}), \qquad \beta^R_{\w'\w} = -(\phi_{\w'},\chi^{R*}_{\w}), \ee
defined by the flat-space scalar product, in spacetime coordinates, 
\be (\phi_{\w'} , \chi^R_{\w} ) = i \int_{0}^{\infty} dx\; \phi_{\w'}^* \stackrel{\leftrightarrow}{\partial_t}\chi^R_{\w}, \ee
or in null coordinates,
\be (\phi_{\w'} , \chi^R_{\w} ) = i \int_{-\infty}^{0} dU\; \phi_{\w'}^* \stackrel{\leftrightarrow}{\partial_U} \chi^R_{\w} \ee
Expanding $(\chi^R_{\w},\chi^R_{\w'})$ in terms of $\phi_{\w'}$ using the transformation \eqn{chiphi2} and utilizing the orthonormality of the $\phi_{\w'}$ modes, the R-completeness relation holds 
\be \label{Rcomplete} \int_0^\infty d\w'' \left[\alpha^R_{\w \w''} \alpha^{R*}_{\w'\w''} - \beta^R_{\w \w''}\beta^{R*}_{\w'\w''}\right] = \delta(\w - \w'). \ee
This is contingent on the orthonormality of both sets of distinct modes by the conserved inner product,
\be \label{phinormal2}(\phi_{\w},\phi_{\w'}) = -(\phi^*_{\w},\phi^*_{\w'}) = \delta(\w-\w'), \ee
\be \label{chinormal2}(\chi^R_{\w},\chi^R_{\w'}) = -(\chi^{R*}_{\w},\chi^{R*}_{\w'}) = \delta(\w-\w'). \ee
and where $(\phi_{\w},\phi^*_{\w'}) = (\chi^R_{\w},\chi^{R*}_{\w'}) = 0$.
As we have seen in a complementary fashion in Section \ref{sec:AIM}, the possible commutation relations are
\be \label{cr1}[a_{\omega}, a_{\omega'}^\dagger]_{\pm} = \int_0^\infty d\w'' \left[\alpha^R_{\w \w''} \alpha^{R*}_{\w'\w''} \pm \beta^R_{\w \w''}\beta^{R*}_{\w'\w''}\right], \ee
\be \label{cr2}[b^R_{\omega}, b^{R \dagger}_{\omega'}]_{\pm} = \int_0^\infty d\w'' \left[\alpha^R_{\w \w''} \alpha^{R*}_{\w'\w''} \pm \beta^R_{\w \w''}\beta^{R*}_{\w'\w''}\right]. \ee
The Wronskian R-completeness relation \eqn{Rcomplete} from the acceleration dynamics connects to the spin-statistics by necessitating the use of the negative sign in the commutation relations \eqn{cr1} and \eqn{cr2}.  It is dictated that
\be [a_\w , a^\dagger_{\w'} ]_{-} = [b^R_\w , b^{R\dagger}_{\w'} ]_{-} = \delta(\w-\w') \ee
and similarly, the zero commuting relations stem from the analogous identities to \eqn{Rcomplete}.  Therefore, in totality, the accelerated Rindler observer entails the spin-statistics connection.

\section{ Conclusions }

The crux of these results are dependent on the presence of acceleration. In the cases presented, acceleration arises as an asymptotically inertial mirror, an infinitely accelerated mirror and the constant acceleration Unruh effect.  That is, in the far future these examples correspond to a proper acceleration $\alpha \rightarrow 0, \infty, \kappa$, respectively.  These new results where found:  1) The assumption of particles at early times obeying late time statistics is un-necessary, as only commutation relations are acceptable in either region, as shown in the asymptotically inertial mirror case. 2) In a non-trivial system, the presence of external acceleration conditions is enough to provide the link to commutation relations. This is in contrast to the specification to dynamic curved spacetime.  Flat-spacetime acceleration conditions allowing non-trivial Bogoliubov transformations establishes the spin-statistics connection. 3) With regards to thermal emission, the time-dependent expression for acceleration in the moving mirror case was explicitly revealed.  This clarifies a salient difference between the Unruh effect and the dynamical Casimir effect.

An acceleration in these above situations gives rise to distinct eigenmode solutions that can be used to represent the field.  One set of eigenmodes can be expressed in mixed positive and negative frequencies pieces of the other set of eigenmodes.  The spin-statistics connection is possible to obtain because of the properties of the conserved inner product and of the existence of these sets of eigenmodes in the first place.  If there was zero acceleration, the Bogoliubov transformation would be trivial and the distinct sets of eigenmodes would not exist.  

\begin{acknowledgments}

MRRG is grateful for discussions with Paul R. Anderson, Charles R. Evans and Xiong Chi.  MRRG appreciates the financial support and hospitality of the IAS-Singapore. 


\end{acknowledgments}

\appendix

\bibliographystyle{unsrt}   
\bibliography{mybib}  

\begin{thebibliography}{10}

\bibitem{Parker:1969au}
Leonard Parker.
\newblock {Quantized fields and particle creation in expanding universes. 1.}
\newblock {\em Phys.Rev.}, 183:1057--1068, 1969.

\bibitem{Parker2}
Leonard Parker.
\newblock Quantized fields and particle creation in expanding universes. ii.
\newblock {\em Phys. Rev. D}, 3:346--356, Jan 1971.

\bibitem{statdyn}
Leonard Parker and Yi~Wang.
\newblock Statistics from dynamics in curved spacetime.
\newblock {\em Phys. Rev. D}, 39:3596--3605, Jun 1989.

\bibitem{nogenstat}
J.~W. Goodison and D.~J. Toms.
\newblock No generalized statistics from dynamics in curved spacetime.
\newblock {\em Phys. Rev. Lett.}, 71:3240--3242, Nov 1993.

\bibitem{Bardek:1994wh}
V.~Bardek, S.~Meljanac, and A.~Perica.
\newblock {Generalized statistics and dynamics in curved space-time}.
\newblock {\em Phys.Lett.}, B338:20--22, 1994.

\bibitem{Goodison:1995sk}
J.W. Goodison.
\newblock {Calogero-Vasilev oscillator in dynamically evolving curved
  space-time}.
\newblock {\em Phys.Lett.}, B350:17--21, 1995.

\bibitem{Parkerghost}
Atsushi Higuchi, Leonard Parker, and Yi~Wang.
\newblock Consistency of faddeev-popov ghost statistics with gravitationally
  induced pair creation.
\newblock {\em Phys. Rev. D}, 42:4078--4081, Dec 1990.

\bibitem{Parkerbook}
Leonard~E Parker and David~J Toms.
\newblock {\em Quantum field theory in curved spacetime: quantized fields and
  gravity}.
\newblock Cambridge monographs on mathematical physics. Cambridge Univ. Press,
  New York, NY, 2009.

\bibitem{BOOYAA}
C.M. et~al. Wilson.
\newblock {Observation of the Dynamical Casimir Effect in a Superconducting
  Circuit, arXiv:1105.4714v1[quant-ph]}.
\newblock 2011.

\bibitem{Wald:1979wt}
Robert~M. Wald.
\newblock {Existence of the S Matrix in Quantum Field Theory In Curved
  Space-time}.
\newblock {\em Annals Phys.}, 118:490--510, 1979.

\bibitem{Davies:1976hi}
P.~C.~W. Davies and S.~A. Fulling.
\newblock {Radiation from a moving mirror in two-dimensional space- time
  conformal anomaly}.
\newblock {\em Proc. Roy. Soc. Lond.}, A348:393--414, 1976.

\bibitem{Davies:1977yv}
P.~C.~W. Davies and S.~A. Fulling.
\newblock {Radiation from Moving Mirrors and from Black Holes}.
\newblock {\em Proc. Roy. Soc. Lond.}, A356:237, 1977.

\bibitem{Carlitz:1986nh}
Robert~D. Carlitz and Raymond~S. Willey.
\newblock {Reflections on Moving Mirrors}.
\newblock {\em Phys. Rev.}, D36:2327, 1987.

\bibitem{Brout:1995rd}
R.~Brout, S.~Massar, R.~Parentani, and Ph. Spindel.
\newblock {A Primer for black hole quantum physics}.
\newblock {\em Phys.Rept.}, 260:329--454, 1995.

\end{thebibliography}

\end{document}